\documentclass[runningheads,a4paper]{llncs}

\usepackage{amssymb}
\usepackage{graphicx}
\usepackage{listings}
\usepackage{amsmath}
\usepackage{fancyvrb}
\usepackage{url}
\urldef{\mailsa}\path|muhammad.khan@dk-compmath.jku.at|
\urldef{\mailsb}\path|Wolfgang.Schreiner@risc.jku.at|

\begin{document}

\mainmatter  

\title{Towards the Formal Specification and Verification of Maple Programs\thanks{The research was funded by the Austrian Science Fund (FWF): W1214-N15, project DK10.}}

\titlerunning{Towards the Formal Specification and Verification of Maple Programs}

%
%
\author{Muhammad Taimoor Khan%
\and Wolfgang Schreiner}
\authorrunning{Muhammad Taimoor Khan and Wolfgang Schreiner}

\institute{Doktoratskolleg Computational Mathematics\\
and\\
Research Institute for Symbolic Computation\\
Johannes Kepler University\\
Linz, Austria\\
\mailsa\\
\mailsb\\
\url{http://www.risc.jku.at/people/mtkhan/dk10/}}

%
%

\maketitle

\enlargethispage*{2cm}
\begin{abstract}
In this paper, we present our ongoing work and initial results on the formal specification and verification of \emph{MiniMaple} (a substantial subset of Maple with slight extensions) programs. The main goal of our work is to find behavioral errors in such programs w.r.t. their specifications by static analysis. This task is more complex for widely used computer algebra languages like Maple as these are fundamentally different from classical languages: they support non-standard types of objects such as symbols, unevaluated expressions and polynomials and require
abstract computer algebraic concepts and objects such as rings and orderings etc. 
As a starting point we have defined and formalized a syntax, semantics, type system and specification language for \emph{MiniMaple}. 

\end{abstract}

\section{Introduction}
Computer algebra programs written in symbolic computation languages such as Maple and Mathematica sometimes do not behave as expected, e.g. by triggering runtime errors or delivering wrong results. There has been a lot of research on applying formal techniques to classical programming languages, e.g. Java~\cite{JML}, C\#~\cite{Spec} and C~\cite{ASCL}; we aim to apply similar techniques to computer algebra languages, i.e. to design and develop a tool for the static analysis of computer algebra programs. This tool will find errors in programs annotated with extra information such as variable types and method contracts, in particular type inconsistencies and violations of methods preconditions.

As a starting point, we have defined the syntax and semantics of a subset of the computer algebra language Maple called \emph{MiniMaple}. Since type safety is a prerequisite of program correctness, we have formalized a type system for \emph{MiniMaple} and implemented a corresponding type checker. The type checker has been applied to the Maple package \emph{DifferenceDifferential}~\cite{ChrDon09} developed at our institute for the computation of bivariate difference-differential dimension polynomials. Furthermore, we have defined a specification language to formally specify the behavior of \emph{MiniMaple} programs. As the next step, we will develop a verification calculus for \emph{MiniMaple} respectively a corresponding tool to automatically detect errors in \emph{MiniMaple} programs with respect to their specifications. An example-based short demonstration of this work is presented in the accompanying paper~\cite{MTK12b}.

Figure 1 gives a sketch of the envisioned system (the verifier component is under development); any \emph{MiniMaple} program is parsed to generate an abstract syntax tree (AST). The AST is then annotated by type information and used by the verifier to check the correctness of a program. Error and information messages are generated by the respective components.

\begin{figure}
  \centering
    \includegraphics[height=7cm,width=5cm]{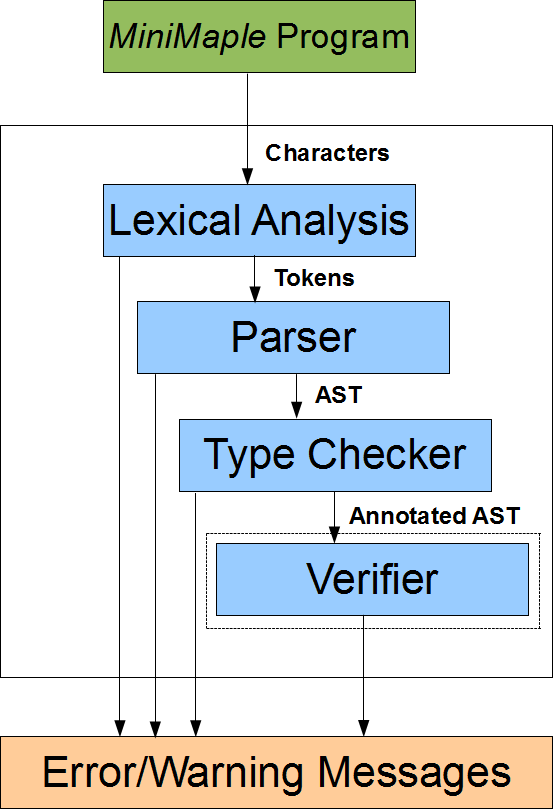}
  \caption{Sketch of the System}
\end{figure}

There are various computer algebra languages, Mathematica and Maple being the most widely used by far, both of which are dynamically typed. We have chosen for our work Maple because of its simpler, more classical and imperative structure.
Still we expect that the results we derive for type checking respective formal specification Maple can be applied to Mathematica, as Mathematica shares with Maple many concepts such as basic kinds of runtime objects.

During our study, we found the following special features for type checking respective formal specification of Maple programs (which are typical for most computer algebra languages):
\begin{itemize}
  \item The language supports some non-standard types of objects, e.g. symbols, unevaluated expressions and polynomials.
  \item The language uses type information to direct the flow of control in the program, i.e. it allows some runtime type-checks which selects the respective code-block for further execution.
  \item The language lacks in the use of abstract data types, which are necessary for the adequate specification of computer algebra functions.
\end{itemize}

The rest of the paper is organized as follows: in Section 2, we describe state of the art related to our work. In Section 3, we introduce the syntax of \emph{MiniMaple} by an example. In Section 4, we briefly explain our type system for \emph{MiniMaple}. In Section 5, we discuss our formal specification language for \emph{MiniMaple}. In Section 6, we highlight the interesting features of a formal semantics of \emph{MiniMaple}. Section 7 presents conclusions and future work.

\enlargethispage*{2cm}
\section{State of the Art}
In this section we first sketch state of the art of type systems for Maple and then discuss the application of formal techniques to computer algebra languages.

Although there is no complete static type system for Maple; there have been several approaches to exploit the type information in Maple for various purposes. For instance, the Maple package Gauss~\cite{Monagan93gauss} introduced parameterized types in Maple. Gauss ran on top of Maple and allowed to implement generic algorithms in Maple in an AXIOM-like manner. The system supported parameterized types and parameterized abstract types, however these were only checked at runtime. The package was introduced in Maple V Release 2 and later evolved into the \emph{domains} package. In~\cite{Carette:2007}, partial evaluation is applied to Maple. The focus of the work is to exploit the available type information for generating specialized programs from generic Maple programs. The language of the partial evaluator has similar syntactic constructs (but fewer expressions) as \emph{MiniMaple} and supports very limited types e.g. integers, rationals, floats and strings. The problem of statically type-checking \emph{MiniMaple} programs is related to the problem of statically type-checking scripting languages such as Ruby~\cite{Furr09}, but there are also fundamental differences due to the different language paradigms.

In comparison to the approaches discussed above, \emph{Mini\-Maple} uses the type annotations provided by Maple for static analysis. It supports a substantial subset of Maple types in addition to named types.

Various specification languages have been defined to formally specify the behavior of programs written in standard classical programming languages, e.g. Java Modeling Language (JML)~\cite{JML} for Java, Spec\#~\cite{Spec} for C\# and ACSL~\cite{ASCL} for ANSI C: these specification languages are used by various tools for extended static checking and verification~\cite{VDG08} of programs written in the corresponding languages. Also variously the application of formal methods to computer algebra has been investigated. For example~\cite{DunLWF} applied the formal specification language Larch~\cite{Gut93} to the computer algebra system AXIOM respective its programming language Aldor. A methodology for Aldor program analysis and verification was devised by defining abstract specifications for AXIOM primitives and then providing an interface between these specifications and Aldor code. The project FoCaLiZe~\cite{Focalize99} aims to provide a programming environment for computer algebra to develop certified programs to achieve high levels of software security. The environment is based on functional programming language FoCal, which also supports some object-oriented features and allows the programmer to write formal specifications and proofs of programs. The work presented in~\cite{Carette:2006} aims at finding a mathematical description of the interfaces between Maple routines. The paper mainly presents the study of the actual contracts in use by Maple routines. The contracts are statements with certain (static and dynamic) logical properties. The work focused to collect requirements for the pure type inference engine for existing Maple routines. The work was extended to develop the partial evaluator for Maple mentioned above~\cite{Carette:2007}. 

The specification language for \emph{MiniMaple} fundamentally differs from those for classical languages such that it supports some non-standard types of objects, e.g. symbols, unevaluated expressions and polynomials etc. The language also supports abstract data types to formalize abstract mathematical concepts, while the existing specification languages are weaker in such specifications. In contrast to the computer algebra specification languages above, our specification language is defined for the commercially supported language Maple, which is widely used but was not designed to support static analysis (type checking respectively verification). The challenge here is to overcome those particularities of the language that hinder static analysis.

\enlargethispage*{2.5cm}
\section{\emph{MiniMaple}}

\emph{MiniMaple} is a simple but substantial subset of Maple that covers all the syntactic domains of Maple but has fewer alternatives in each domain than Maple; in particular, Maple has many expressions which are not supported in our language. The complete syntactic definition of \emph{MiniMaple} is given in \cite{MTK11a}. The grammar of \emph{MiniMaple} has been formally specified in BNF from which a parser for the language has been automatically generated with the help of the parser generator ANTLR.

The top level syntax for \emph{MiniMaple} is as follows:
\begin{quote}
 \begin{tabbing}
\emph{Prog} := \emph{Cseq}; \\
\emph{Cseq} := EMPTY $|$ \emph{C,Cseq}\\
\emph{C} := ... $|$ \emph{I,Iseq} := \emph{E,Eseq} $|$ ...
\end{tabbing}
\end{quote}

A program is a sequence of commands, there is no separation between declaration and assignment.
\begin{footnotesize}
\begin{tabbing} \emph{1}. status:=0; \\\emph{2}. prod := \= \textbf{proc}(l::\textbf{list}(\textbf{Or}(\textbf{integer},\textbf{float})))::[\textbf{integer},\textbf{float}];
\\\emph{3}. \> \textbf{global} status; 
\\\emph{4}. \> \textbf{local} i::\textbf{integer}, x::\textbf{Or}(\textbf{integer},\textbf{float}), si::\textbf{integer}:=1, sf::\textbf{float}:=1.0; 
\\\emph{5}. \> \textbf{for} \= i \textbf{from} 1 \textbf{by} 1 \textbf{to} nops(l) \textbf{while} (running) \textbf{do}
\\\emph{6}. \> \> x:=l[i];
\\\emph{7}. \> \> \textbf{if} \textbf{type}(\= x,integer) \textbf{then}
\\\emph{8}. \> \> \> \textbf{if} (\=x = 0) \textbf{then}
\\\emph{9}. \> \> \> \> return [si,sf];
\\\emph{10}. \> \> \> \textbf{else}
\\\emph{11}. \> \> \> si:=si*x;
\\\emph{12}. \> \> \> \textbf{end if};
\\\emph{13}. \> \> \textbf{elif} \textbf{type}\=(x,float) \textbf{then}
\\\emph{14}. \> \> \> \textbf{if} (\=x $<$ 0.5) \textbf{then}
\\\emph{15}. \> \> \> \> return [si,sf];
\\\emph{16}. \> \> \> \textbf{else}
\\\emph{17}. \> \> \> sf:=sf*x;
\\\emph{18}. \> \> \> \textbf{end if};
\\\emph{19}. \> \>\textbf{end if};
\\\emph{20}. \> \textbf{end do;}
\\\emph{26}. \> \textbf{return} [si,sf];
\\\emph{27}. \> \textbf{end proc};
\end{tabbing}
\end{footnotesize}
\begin{center}
 Listing 1: An example \emph{MiniMaple} program
\end{center}
Listing 1 gives an example of a \emph{MiniMaple} program which we will use in the following sections for the discussion of type checking and behavioral specification. The program consists of an assignment initializing a global variable \emph{status} and an assignment defining a procedure \emph{prod} followed by the application of the procedure. The procedure takes a list of integers and floats and computes the product of these integers and floats separately; it returns as a result a tuple of the products. The procedure may also terminate prematurely for certain inputs, i.e. either for an integer value 0 or for a float value less than $0.5$ in the list; in this case the procedure computes the respective products just before the index at which the aforementioned terminating input occurs. 


As one can see from the example, we make use of the type annotations that Maple introduced for runtime type checking. In particular, we demand that function parameters, function results and local variables are correspondingly type annotated. Based on these annotations, we define a language of types and a corresponding type system for the static type checking of \emph{MiniMaple} programs.


\enlargethispage*{2cm}
\section{A Type System for \emph{MiniMaple}}
A \emph{type} is (an upper bound on) the range of values of a variable. A \emph{type system} is a set of formal typing rules to determine the variables types from the text of a program. A type system prevents \emph{forbidden errors} during the execution of a program. It completely prevents the \emph{untrapped errors} and also a large class of \emph{trapped errors}. \emph{Untrapped errors} may go unnoticed for a while and later cause an arbitrary behavior during execution of a program, while \emph{trapped errors} immediately stop execution~\cite{Cardelli97}.

A type system in essence is a decidable logic with various kinds of \emph{judgments}; for example the typing judgment 
\begin{center}
$\pi$ $\vdash$ \emph{E}:($\tau$)\textbf{exp} 
\end{center}
can be read as ``in the given type environment $\pi$, \emph{E} is a well-typed expression of type $\tau$''. A type system is \emph{sound}, if the deduced types indeed capture the program values exhibited at runtime. 

In the following we describe the main properties of a type system for \emph{Mini\-Maple}. Subsection 4.1 sketches its design and Subsection 4.2 presents its implementation and application. A proof of the soundness of the type system remains to be performed.

\enlargethispage*{2cm}
\subsection{Design}
\emph{MiniMaple} uses Maple type annotations for static type checking, which gives rise to the following language of types:
\begin{tabbing}
\emph{T} ::= \= \textbf{integer} $|$ \textbf{boolean} $|$ \textbf{string} $|$ \textbf{float} $|$ \textbf{rational} $|$ \textbf{anything}\\\> $|$ \textbf{\{} \emph{T} \textbf{\}} $|$ \textbf{list(} \emph{T} \textbf{)} $|$ \textbf{[} \emph{Tseq} \textbf{]} $|$ \textbf{procedure[} \emph{T} \textbf{](} \emph{Tseq} \textbf{)} \\
\> $|$ \emph{I}( \emph{Tseq} ) $|$ \textbf{Or(} \emph{Tseq} \textbf{)} $|$ \textbf{symbol} $|$ \textbf{void} $|$ \textbf{uneval} $|$ \emph{I} 
\end{tabbing}

The language supports the usual concrete data types, sets of values of type \emph{T} (\{ \emph{T} \}), lists of values of type \emph{T} (\textbf{list}( \emph{T} )) and records whose members have the values of types denoted by a type sequence \emph{Tseq} ([ \emph{Tseq} ]). Type \textbf{anything} is the super-type of all types. Type \textbf{Or}( \emph{Tseq} ) denotes the union type of various types, type \textbf{uneval} denotes the values of unevaluated expressions, e.g. polynomials, and type \textbf{symbol} is a name that stands for itself if no value has been assigned to it. User-defined data types are referred by \emph{I} while \emph{I}( \emph{Tseq} ) denotes tuples (of values of types \emph{Tseq}) tagged by a name \emph{I}.

A sub-typing relation ($<$) is defined among types, i.e. \textbf{integer} $<$ \textbf{rational} $<$ ... $<$ \textbf{anything}, such that \textbf{integer} is a sub-type of \textbf{rational} and the type \textbf{anything} is the super-type of all types.

In the following, we demonstrate the problems arising from type checking \emph{MiniMaple} programs using the example presented in the previous section.
\subsubsection*{Global Variables}
Global variables (declarations) can not be type annotated; therefore to global variables values of arbitrary types can be assigned in Maple. We introduce \emph{global} and \emph{local} contexts to handle the different semantics of the variables inside and outside of the body of a procedure respective loop. 
 \begin{itemize}
  \item In a \emph{global} context new variables may be introduced by assignments and the types of variables may change arbitrarily by assignments.
 \item In a \emph{local} context variables can only be introduced by declarations. The types of variables can only be \emph{specialized} i.e. the new value of a variable should be a sub-type of the declared variable type. The sub-typing relation is observed while specializing the types of variables.
 \end{itemize}
\subsubsection*{Type Tests}
A predicate \textbf{type}(\emph{E,T}) (which is true if the value of expression $E$ has type $T$) may direct the control flow of a program. If this predicate is used in a conditional, then different branches of the conditional may have different type information for the same variable. We keep track of the type information introduced by the different type tests from different branches to adequately reason about the possible types of a variable. For instance, if a variable x has type Or(integer,float), in a conditional statement where the "then" branch is guarded by a test type(x,integer), in the "else" branch x has automatically type float. A warning is generated, if a test is redundant (always yields true or false).

\enlargethispage*{4cm}
\begin{footnotesize}
\begin{tabbing} \emph{1}. status:=0; \\ \emph{2}. prod := \= \textbf{proc}(l::\textbf{list}(\textbf{Or}(\textbf{integer},\textbf{float})))::[\textbf{integer},\textbf{float}];
\\\emph{3}. \> \textsl{\# $\pi$=\{l:\textbf{list}(\textbf{Or}(\textbf{integer},\textbf{float}))\}}
\\\emph{4}. \> \textbf{global} status; 
\\\emph{5}. \> \textbf{local} i, x::\textbf{Or}(\textbf{integer},\textbf{float}), si::\textbf{integer}:=1, sf::\textbf{float}:=1.0;
\\\emph{6}. \> \textsl{\# $\pi$=\{..., i:\textbf{symbol}, x:\textbf{Or}(\textbf{integer},\textbf{float}),..., status:\textbf{anything}\}}
\\\emph{7}. \> \textbf{for} \= i \textbf{from} 1 \textbf{by} 1 \textbf{to} nops(l) \textbf{do}
\\\emph{8}. \> \> x:=l[i]; status:=i;
\\\emph{10}. \> \> \textsl{\# $\pi$=\{..., i:\textbf{integer}, ..., status:\textbf{integer}\}}
\\\emph{11}. \> \> \textbf{if} \textbf{type}(\= x,integer) \textbf{then}
\\\emph{12}. \> \> \textsl{\# $\pi$=\{..., i:\textbf{integer}, x:\textbf{integer}, si:\textbf{integer}, ..., status:\textbf{integer}\}}
\\\emph{13}. \> \> \> \textbf{if} (\=x = 0) \textbf{then} \textbf{return} [si,sf]; \textbf{end if};
\\\emph{16}. \> \> \> si:=si*x;
\\\emph{17}. \> \> \textbf{elif} \textbf{type\=}(x,float) \textbf{then}
\\\emph{18}. \> \> \textsl{\# $\pi$=\{..., i:\textbf{integer}, x:\textbf{float}, ..., sf:\textbf{float}, status:\textbf{integer}\}}
\\\emph{19}. \> \> \> \textbf{if} (\=x $<$ 0.5) \textbf{then} \textbf{return} [si,sf]; \textbf{end if};
\\\emph{22}. \> \> \> sf:=sf*x;
\\\emph{23}. \> \>\textbf{end if};
\\\emph{24}. \> \> \textsl{\# $\pi$=\{..., i:\textbf{integer}, x:\textbf{Or}(\textbf{integer},\textbf{float}),..., status:\textbf{integer}\}}
\\\emph{25}. \> \textbf{end do;}
\\\emph{26}. \> \textsl{\# $\pi$=\{..., i:\textbf{symbol}, x:\textbf{Or}(\textbf{integer},\textbf{float}),..., status:\textbf{anything}\}}
\\\emph{27}. \> status:=-1;
\\\emph{28}. \> \textsl{\# $\pi$=\{..., i:\textbf{symbol}, x:\textbf{Or}(\textbf{integer},\textbf{float}),..., status:\textbf{integer}\}}
\\\emph{29}. \> \textbf{return} [si,sf];
\\\emph{30}. \> \textbf{end proc};
\\\emph{31}. result := prod([1, 8.54, 34.4, 6, 8.1, 10, 12, 5.4]);
\end{tabbing}
\end{footnotesize}
\begin{center}
 Listing 2: A \emph{MiniMaple} procedure type-checked
\end{center}
For our example program our type system will generate the type information as depicted in Listing 2. The program is annotated with the type environment (a partial function from identifiers to their corresponding types) of the form \#$\pi = $\{\emph{variable}:\emph{type},...\}. For example, the type environment at line \emph{6} shows the types of the respective variables as determined by the static analysis of parameter and identifier declarations (\textbf{global} and \textbf{local}).

The static analysis of the two branches of the conditional command in the body of the loop introduces the type environments at lines \emph{12} and \emph{18} respectively; the type of variable \emph{x} is determined as \textbf{integer} and \textbf{float} by the conditional type-expressions respectively. 

There is more type information to direct the program control flow for an identifier \emph{x} introduced by an expression \textbf{type}(\emph{E,T}) at lines \emph{11} and \emph{17}.

By analyzing the conditional command as a whole, the type of variable \emph{x} is determined as \textbf{Or(integer, float)} (at line \emph{24}), i.e. the union type of the two types determined by the respective branches.

The local type information introduced/modified by the analysis of body of loop does not effect the global type information. The type environment at lines \emph{6} and \emph{26} reflects this fact for variables \emph{status}, \emph{i} and \emph{x}. This is because of the fact that the number of loop iterations might have an effect on the type of the variable otherwise and one cannot determine the concrete type by the static analysis. To handle this non-determination of types we put a reasonable upper bound (fixed point) on the types of such variables, namely the type of a variable prior to the body of a loop. 

%
%

\subsection{Formalization}
In this subsection we explain the typing judgments and typing rules for some expressions and commands of \emph{MiniMaple}. These judgments use the following kinds of objects (``Identifier'' and ''Type`` are the syntactic domains of identifiers/variables and types of \emph{MiniMaple} respectively):
\begin{itemize}
  \item $\pi$: Identifier $\rightarrow$ Type: a type environment, i.e. a (partial) function from identifiers to types.
  \item \emph{c} $\in$ \{global, local\}: a tag representing the context to check if the corresponding syntactic phrase is type checked inside/outside of the procedure/loop.
  \item \emph{asgnset} $\subseteq$ Identifier: a set of assignable identifiers introduced by type checking the declarations.
  \item $\epsilon set$ $\subseteq$ Identifier: a set of thrown exceptions introduced by type checking the corresponding syntactic phrase.
  \item $\tau set$ $\subseteq$ Type: a set of return types introduced by type checking the corresponding syntactic phrase.
  \item \emph{rflag} $\in$ \{aret, not\_aret\}: a return flag to check if the last statement of every execution of the corresponding syntactic phrase is a \emph{return} command.
\end{itemize}
\enlargethispage*{2cm}
\emph{MiniMaple} supports various types of expressions but boolean expressions are treated specially because of the test \textbf{type}(\emph{I,T}) that gives additional type information about the expression.
The typing judgment for boolean expressions
\begin{center}
$\pi$ $\vdash$ \emph{E}:($\pi_1$)\textbf{boolexp}
\end{center}
can be read as ''with the given $\pi$, \emph{E} is a well-typed boolean expression with new type environment $\pi_1$``. The new type environment is produced as a fact of type test that might introduce new type information for an identifier.

The typing judgment for commands
\begin{center}
$\pi$, \emph{c}, \emph{asgnset} $\vdash$ \emph{C}:($\pi_1$, $\tau set$, $\epsilon set$, \emph{rflag})\textbf{comm}
\end{center}
can be read as "in the given type environment $\pi$, context \emph{c} and an assignable set of identifiers \emph{asgnset}, \emph{C} is a well-typed command and produces ($\pi_1$, $\tau set$, $\epsilon set$, \emph{rflag}) as type information''.
\enlargethispage*{2cm}
In the following we explain some typing rules to derive typing judgments for boolean expressions and conditional commands. These typing rules use different kinds of auxiliary functions and predicates as given below.
\subsubsection*{Auxiliary Functions}
\begin{itemize}
  \item \emph{specialize}$(\pi_1,\pi_2)$: specializes the identifiers of former type environment to the identifiers in the latter type environment w.r.t. their types.
  \item \emph{combine}$(\pi_1,\pi_2)$: combines the identifiers in the two environments with respect to their types.
  \item \emph{superType}$(\tau_1,\tau_2)$: returns the super-type between the two given types.
\end{itemize}
\subsubsection*{Auxiliary Predicates}
\begin{itemize}
  \item \emph{canSpecialize}$(\pi_1,\pi_2)$: returns true if all the common identifiers (in both type environments) have a super-type between their corresponding types.
  \item \emph{superType}$(\tau_1,\tau_2)$: returns true (in most cases) if the former type is general (super) type than the latter type. \textbf{Anything} is the super-type of all types.
\end{itemize}
\subsubsection*{Typing Rules}
The typing rule for boolean expressions is as follows:
\begin{itemize}
 \item \textbf{type}(\emph{I,T})
\begin{center}
\underline{$\pi$ $\vdash$ \emph{I}:($\tau_1$)\textbf{id} \hspace*{0.25cm} $\pi$ $\vdash$ \emph{T}:($\tau_2$)\textbf{type} \hspace*{0.25cm} \emph{superType}($\tau_1$,$\tau_2$)}
\\ $\pi$ $\vdash$ \textbf{type}(\emph{I,T}):(\{I:$\tau_2$\})\textbf{boolexp}
\end{center}
The phrase ``\textbf{type}(\emph{I,T})`` is a well-typed boolean expression if the declared type of identifier ($\tau_1$) is the super-type of \emph{T} ($\tau$). The boolean expression may introduce new type information for the identifier.
\end{itemize}

The typing rule for the conditional command is given below:
\begin{itemize}

\item \textbf{if} \emph{E} \textbf{then} \emph{Cseq} \emph{Elif} \textbf{end if}
\begin{center}
$\pi$ $\vdash$ \emph{E}: ($\pi$')\textbf{boolexp} \quad \emph{canSpecialize}($\pi$,$\pi$') \\ \emph{specialize}($\pi$,$\pi$'), \emph{c}, \emph{asgnset} $\vdash$ \emph{Cseq}:($\pi_1$,$\tau set_1$,$\epsilon set_1$,$rflag_1$)\textbf{cseq} \\ \underline{$\pi$, \emph{c}, \emph{asgnset} $\vdash$ \emph{Elif}:($\pi_2$,$\pi set$, $\tau set_2$,$\epsilon set_2$,$rflag_2$)\textbf{elif}}
\\$\pi$, \emph{c}, \emph{asgnset} $\vdash$ \textbf{if} \emph{E} \textbf{then} \emph{Cseq} Elif \textbf{end if}:(\emph{combine}($\pi_1$,$\pi_2$),$\tau set_1 \cup \tau set_2$,$\epsilon set_1 \cup \epsilon set_2$,ret($rflag_1$, $rflag_2$))\textbf{comm}
\end{center}
The phrase ``\textbf{if} \emph{E} \textbf{then} \emph{Cseq} \emph{Elif} \textbf{end if}`` is a well typed conditional command if the type of expression \emph{E} does not conflict global type information. The conditional command combines the type environment of its two conditional branches (\emph{if} and \emph{elif}), because we are not sure which of the branches will be executed at runtime.
\end{itemize}
\subsection{Application}
Based on the type system sketched above we have implemented a type checker for \emph{MiniMaple}~\cite{MTK11a} in Java (150$+$ classes and 15K$+$ lines of code).  The type checker also handles the specification language of \emph{MiniMaple}.

Figure 2 shows that the output of the type checker applied to a file containing the source code of the example program from the previous section. It shows that the file has successfully parsed and also presents the type annotations for the first assignment command. In the second part, it shows the resulting type environment with the associated program identifiers and their respective types introduced while type checking. The last message indicates that the program type checked correctly.

\begin{figure}[!h]
\begin{center}
\begin{scriptsize}
\begin{verbatim}
/home/taimoor/antlr3/Test6.m parsed with no errors.
Generating Annotated AST...
...
**********COMMAND-SEQUENCE-ANNOTATION START**********
PI -> [
prod:procedure[[integer,float]](list(Or(integer,float)))
status:integer
result:[integer,float]
]
RetTypeSet -> {}
ThrownExceptionSet -> {}
RetFlag -> not_aret
**********COMMAND-SEQUENCE-ANNOTATION END************
Annotated AST generated.
The program type-checked correctly.
\end{verbatim}
\end{scriptsize}
\caption{Parsing and Type Checking the Program}
\end{center}
\end{figure}

The main test case for our type checker is the Maple package \emph{Difference\-Differential}~\cite{ChrDon09} developed by Christian D{\"o}nch at our institute. The package provides algorithms for computing difference-differential dimension polynomials by relative Gr{\"o}bner bases in difference-differential modules according to the method developed by M. Zhou and F. Winkler~\cite{Zhou08}.

We manually translated this package into a \emph{MiniMaple} package so that the type checker can be applied. This translation consists of 
\begin{itemize}
 \item adding required type annotations and
  \item translating those parts of the package that are not directly supported into logically equivalent \emph{MiniMaple} constructs.
\end{itemize}
No crucial typing errors have been found but some bad code parts have been identified that can cause problems, e.g., variables that are declared but not used (and therefore cannot be type checked) and variables that have duplicate global and local declarations.
\section{A Formal Specification Language for \emph{MiniMaple}}
Based on the type system presented in the previous section, we have developed a formal specification language for \emph{MiniMaple}. This language is a logical formula language which is based on Maple notations but extended by new concepts. The formula language supports various forms of quantifiers, logical quantifiers (\textbf{exists} and \textbf{forall}), numerical quantifiers (\textbf{add}, \textbf{mul}, \textbf{min} and \textbf{max}) and sequential quantifier (\textbf{seq}) representing truth values, numeric values and sequence of values respectively. We have extended the corresponding Maple syntax, e.g., logical quantifiers use typed variables and numerical quantifiers are equipped with logical conditions that filter values from the specified variable range. The example for these quantifiers is explained later in the procedure specification of this section. The use of this specification language is described in the conclusions.

Also the language allows to formally specify the behavior of procedures by pre- and post-conditions and other constraints; it also supports loop specifications and assertions. In contrast to specification languages such as JML, abstract data types can be introduced to specify abstract concepts and notions from computer algebra.


\enlargethispage*{2cm}
At the top of \emph{MiniMaple} program one can declare respectively define mathematical functions, user-defined named and abstract data types and axioms. The syntax of specification declarations
\begin{quote}
 \begin{tabbing}
\emph{decl} ::= \= EMPTY \\\>$|$ (\=\textbf{define}(\emph{I,rules}); \\\>\>$|$ \textbf{`type/}\emph{I}`:=\emph{T}; $|$ \textbf{`type/}\emph{I}`;\\\>\> $|$ \textbf{assume}(\emph{spec-expr}); ) \emph{decl} 
\end{tabbing} 
\end{quote}
is mainly borrowed from Maple. The phrase ``\textbf{define}(\emph{I,rules});`` can be used for defining mathematical functions as shown in the following the factorial function:
\begin{quote}
 \begin{tabbing}
 \textbf{define}(fac, fac(0) = 1, fac(n::integer) = n * fac(n -1));
\end{tabbing}
\end{quote}
User-defined data types can be declared with the phrase ''\textbf{`type/}\emph{I}`:=\emph{T};`` as shown in the following declaration of ''ListInt'' as the list of integers:
\begin{quote}
 \begin{tabbing}
\textbf{`type/}ListInt`:=\textbf{list}(\textbf{integer});
\end{tabbing}
\end{quote}

The phrase ''\textbf{`type/}\emph{I}`;'' can be used to declare abstract data type with the name \emph{I}, e.g. the following example shows the declaration of abstract data type ``difference differential operator (DDO)''.
\begin{quote}
 \begin{tabbing}
\textbf{`type/}DDO`;
\end{tabbing}
\end{quote}
The task of formally specifying mathematical concepts using abstract data types is more simpler as compared to their underlying representation with concrete data types. Also other related facts and the access functions of abstract concept can be formalized for better reasoning.

Axioms can be introduced by the phrase ``\textbf{assume}(\emph{spec-expr});`` as the following example shows an axiom that an operator is a difference-differential operator, if its each term is a difference-differential term, where d is an operator:
\begin{quote}
 \begin{tabbing}
\textbf{assume}(\=isDDO(d) equivalent forall(i::integer, 1$<=$i and i$<=$terms(d) implies
\\\>isDDOTerm(\=getTerm(d,i,1),getTerm(d,i,2),
\\\>\>getTerm(d,i,3),getTerm(d,i,4));
\end{tabbing}
\end{quote}
Any predicate declaration can be introduced by the phrase ``\emph{I}(\emph{spec-expr});`` as the following example shows a predicate that when a given field is supported:

\begin{center}
 inField(c);
\end{center}

The entities introduced by the specification declarations can be used in the following specifications.

A procedure specification consists of a pre-condition, the set of global variables that can be modified and the post condition, describing the relationship between pre and post state. By an optional exception clause we can specify the exceptional behavior of a procedure. The procedure specification syntax is influenced by the Java Modeling Language:
\begin{quote}
 \begin{tabbing}
\emph{proc-spec} ::= \= \textbf{requires} \emph{spec-expr};\\\> \textbf{global} \emph{Iseq};\\\> \textbf{ensures} \emph{spec-expr}; \emph{excep-clause}
\end{tabbing}
\end{quote}
Listing 3 shows an example for the procedure specification. The specification is a big logical disjunction to formulate two possible behaviors of the procedure:
\begin{enumerate}
  \item when the procedure terminates normally and
   \item when the procedure terminates prematurely.
 \end{enumerate}
\enlargethispage*{1.25cm}
\begin{footnotesize}
(*@
\\ \hspace*{3 mm} \textbf{requires} true;
\\ \hspace*{3 mm} \textbf{global} status;
\\ \hspace*{3 mm} \textbf{ensures}
\\ \hspace*{3 mm} \hspace*{3 mm}(status = -1 and RESULT[1] = mul(e, e in l, type(e,integer))
\\ \hspace*{3 mm} \hspace*{3 mm}and RESULT[2] = mul(e, e in l, type(e,float))
\\ \hspace*{3 mm} \hspace*{3 mm}and forall(i::integer, 1$<=$i and i$<=$nops(l) and type(l[i],integer) implies l[i]$<>$0)
\\ \hspace*{3 mm} \hspace*{3 mm}and forall(i::integer, 1$<=$i and i$<=$nops(l) and type(l[i],float) implies l[i]$>=$0.5))
\\ \hspace*{3 mm} \hspace*{3 mm}or
\\ \hspace*{3 mm} \hspace*{3 mm}(1$<=$status and status$<=$nops(l)
\\ \hspace*{3 mm} \hspace*{3 mm}and RESULT[1] = mul(l[i], i=1..status-1, type(l[i],integer))
\\ \hspace*{3 mm} \hspace*{3 mm}and RESULT[2] = mul(l[i], i=1..status-1, type(l[i],float))
\\ \hspace*{3 mm} \hspace*{3 mm}and ((type(l[status],integer) and l[status]=0)
\\ \hspace*{3 mm} \hspace*{10 mm}or (type(l[status],float) and l[status]$<$0.5))
\\ \hspace*{3 mm} \hspace*{3 mm}and forall(i::integer, 1$<=$i and i$<$status and type(l[i],integer) implies l[i]$<>$0)
\\ \hspace*{3 mm} \hspace*{3 mm}and forall(i::integer, 1$<=$i and i$<$status and type(l[i],float) implies l[i]$>=$0.5));
\\ \hspace*{3 mm} @*)
\\ \textbf{proc}(l::\textbf{list}(\textbf{Or}(\textbf{integer},\textbf{float})))::[\textbf{integer},\textbf{float}]; ... \textbf{end proc};

\end{footnotesize}
\enlargethispage*{2cm}
\begin{center}
 Listing 3: A \emph{MiniMaple} procedure formally specified
\end{center}

The listing gives a formal specification of the example procedure introduced in Section 3. The procedure has no pre-condition as shown in the \textbf{requires} clause; the \textbf{global} clause says that a global variable \emph{status} can be modified by the body of the procedure. The normal behavior of the procedure is specified in the \textbf{ensures} clause. 

The post condition specifies that, if the complete list is processed then we get the result as the product of all integers and floats in the list; if the procedure terminates pre-maturely, then we only get the product of integers and floats till the value of variable \emph{status} (index of the input list).

From the example one can also notice the application of numerical quantifier \textbf{mul}. The quantifier multiplies only those elements of the input array \emph{l} that satisfy the test \textbf{type}(e,\textbf{integer}).

Loops can be specified by invariants and termination terms denoting non-negative integers as follows:
\begin{quote}
 \begin{tabbing}
\emph{loop-spec} := \textbf{invariant} \emph{spec-expr}; \textbf{decreases} \emph{spec-expr};
\end{tabbing}
\end{quote}

The following example specifies the loop that iterates over integers from 1...100 respectively computes the sum.
\begin{quote}
 \begin{tabbing}
i := 1;
s := 0;
n := 100;\\
\textbf{while} (i $<=$ n) \textbf{do}\{\\
(*@\textbf{invariant} s = \textbf{OLD} s + i - 1; \textbf{decreases} n-i;@*)\\
s := s + i; i := i + 1;\\
\}
\end{tabbing}
\end{quote}
From the example one can see the relationship between the loop variables that holds after every iteration and that the value of the termination term decreases after every iteration.

Loop specifications help in reasoning about loops, i.e. about partial correctness (invariants) and total correctness (termination term).

\enlargethispage*{2cm}


Assertions have Maple borrowed syntax as given:
\begin{quote}
 \begin{tabbing}
 \emph{asrt} := \textbf{ASSERT}(\emph{spec-expr}, (EMPTY $|$ ``\emph{I}``));
\end{tabbing}
\end{quote}
An assertion can be a logical formula or a named assertion. The following example shows a named assertion (''test failed'').
\begin{quote}
 \begin{tabbing}
x := 1; y := x;
x := x + y;\\
\textbf{ASSERT}(\textbf{type}(y,\textbf{integer}), ''test failed``);
\end{tabbing}
\end{quote}
The implemented type checker also checks the correct typing of the formal specifications. We have used the specification language to formally specify parts of \emph{DifferenceDifferential}. While the specifications are not yet formally checked, they demonstrate the adequacy of the language for the intended purpose.

\enlargethispage*{2cm}
\section{Formal Semantics of \emph{MiniMaple}}
We have defined a formal denotational semantics of \emph{MiniMaple} programs as a pre-requisite of a verification calculus which we are currently developing: the verification conditions generated by the verification calculus must be sound with respect to the semantics. There is no formally defined semantics for Maple and only the implementation of Maple can be considered as a basis for our work. However, our semantics definition attempts to depict the internal behavior of Maple. Based on this semantics, now we can ask the question about the correct behavior of any \emph{MiniMaple} program. The complete definition of a formal semantics of \emph{Mini\-Maple} is given in~\cite{MTK12a}. Its core features are as follows:

\begin{itemize}
 \item \emph{MiniMaple} has expressions with side-effects, which is not supported in functional programming languages like Haskel~\cite{haskel00} and Miranda~\cite{LambertLR93}. As a result the evaluation of an expression may change the state. The formal semantics of expression evaluation and command execution is therefore defined as a state relationship between pre- and post-states. A formal denotational semantics is defined as a state relationship is easier to integrate with non-uniquely specified procedures as compared to the function-based semantics definition~\cite{WS08}.
 \item Semantic domains of values have some non-standard types of objects, for example symbol, uneval and union etc. \emph{MiniMaple} also supports additional functions and predicates, for example type tests i.e. \textbf{type}(\emph{E,T}), which are correspondingly modeled in semantics algebras.
 \item In \emph{MiniMaple} a procedure is introduced by an assignment command, e.g.\\ I := \textbf{proc}() $\ldots$ \textbf{end proc}, such that assignments take the role of declarations in classical languages. Furthermore, static scoping is used in the definition of a \emph{MiniMaple} procedure.
\end{itemize}

The denotational semantics is based on \emph{semantic algebra}~\cite{Schmidt86}. For example \emph{Value} is a disjunctive union domain composed of all kinds of primitive semantic values (domains) supported in \emph{MiniMaple}. It also defines some interesting domains i.e. \emph{Module}, \emph{Procedure}, \emph{Uneval} and \emph{Symbol}. The domain \emph{Value} is a recursive domain, e.g. \emph{List} is defined by \emph{Value*} as follows:
\begin{quote}
$List = Value*$
\\$Value = ... | List | ...$
\end{quote}

A valuation function defines a mapping of a language's abstract syntax structures to its corresponding meaning (semantic algebras)~\cite{Schmidt86}. A valuation function D for a syntax domain D is usually formalized by a set of equations, one per alternative in the corresponding BNF rule for the \emph{MiniMaple} syntactic domain. As the formal semantics of \emph{MiniMaple} is defined as a state relationship so we define the result of valuation function as a predicate. For example the state relation (\emph{StateRelation}) is defined as a power set of pair of pre- and post-states as follows:
\begin{center}
 \emph{StateRelation} := $\mathbb{P}(State \times StateU)$
\end{center}

The valuation function for command sequence takes the abstract syntax of command sequence, a value of type Cseq and type environment \emph{Environment} and results in a \emph{StateRelation} as follows:
\begin{center}
[[ Cseq ]] : \emph{Environment} $\rightarrow$ \emph{StateRelation}
\end{center}

The denotational semantics of \emph{MiniMaple} while-loop is defined as a relationship between a pre-state $s$ and post-state $s'$ as follows:
\begin{tabbing}
[[ \textbf{while} \= $E$ \textbf{do} $Cseq$ \textbf{end do} ]](e)(s,s') $\Leftrightarrow$ 
\\\> $\exists $ \= $k \in Nat', t, u \in StateU*: t(1)=inStataU(s) \wedge u(1)=inStateU(s) \wedge$
\\\>\> $( \forall i \in Nat'_k: $ \= $iterate(i, t, u, e, [[ E ]], [[ Cseq ]]) ) \wedge$
\\\>\>\> $( ( u(k)=inError() \wedge s'=u(k) ) \vee$
\\\>\>\> $( returns(data(inState(u(k)))) \wedge s'=t(k) ) \vee$
\\\>\>\> $( \exists $ \= $v \in ValueU: [[E]](e)(inState(t(k)), u(k), v)$
\\\>\>\>\> $\wedge v <> inValue(inBoolean(True)) \wedge$ 
\\\>\>\>\> IF \= $v = inValue(inBoolean(False))$ THEN
\\\>\>\>\>\> s'=t(k)
\\\>\>\>\> ELSE $s' = inError()$ END
\\\>\>\>) 
\\\>\>)
\end{tabbing}

The corresponding \emph{iterate} predicate formalizes the aforementioned while-loop semantics. For the complete list of semantic algebras, domains and valuation functions, please see~\cite{MTK12a}.

\enlargethispage*{2.5cm}
\section{Conclusions and Future Work}
In this paper we gave an overview of \emph{Mini\-Maple} and its formal type system. We plan to automatically infer types as a future goal. Also we presented our initial work on a formal specification language for \emph{MiniMaple} that can be used to specify the behavior of \emph{MiniMaple} programs. As a main test case we have used our specification language to formally specify various abstract computer algebraic concepts used in the Maple package \emph{DifferenceDifferential}, e.g. difference-differential operator and terms and various related access functions. We may use this specification language to generate executable assertions that are embedded in \emph{Mini\-Maple} programs and check at runtime the validity of pre/post conditions. Our main goal, however, is to use the specification language for static analysis, in particular to detect violations of methods preconditions. For this purpose, based on the results of a prior investigation we intend to use the verification framework Why3~\cite{Why3} to implement the verification calculus for \emph{MiniMaple} as depicted in Fig. 3.

\begin{figure}
  \centering
     \includegraphics[scale=0.65]{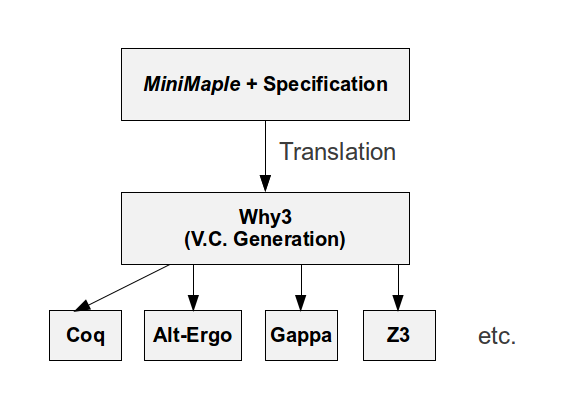}
  \caption{Verification Calculus for \emph{MiniMaple}}
\end{figure}

As one can see in the figure, here we need to translate our specification-annotated \emph{MiniMaple} program into the intermediate language of Why3 and then use the various proving back-ends of Why3. Currently we are working on this translation.



%

\enlargethispage*{2cm}
\bibliography{paper}
\bibliographystyle{plain}

\end{document}